\documentclass[sigconf,natbib=true,anonymous=false]{acmart}
\usepackage[utf8]{inputenc}
\usepackage[shortlabels]{enumitem}
\usepackage{listings}
\usepackage{float}
\usepackage{tabularx}
\usepackage{booktabs}       
\usepackage{multirow}

\renewcommand\footnotetextcopyrightpermission[1]{} 
\begin{document}

\title{InPars Toolkit: A Unified and Reproducible Synthetic Data Generation Pipeline for Neural Information Retrieval}
\author{Hugo Abonizio}
\affiliation{%
  \institution{NeuralMind}
  \institution{University of Campinas}
  \country{Brazil}
}
\author{Luiz Bonifacio}
\affiliation{%
  \institution{NeuralMind}
  \institution{University of Campinas}
  \country{Brazil}
}
\author{Vitor Jeronymo}
\affiliation{%
  \institution{NeuralMind}
  \institution{University of Campinas}
  \country{Brazil}
}
\author{Roberto Lotufo}
\affiliation{%
  \institution{NeuralMind}
  \institution{University of Campinas}
  \country{Brazil}
}
\author{Jakub Zavrel}
\affiliation{%
    \institution{Zeta Alpha}
    \country{Netherlands}
}
\author{Rodrigo Nogueira}
\affiliation{%
  \institution{Zeta Alpha}
  \institution{NeuralMind}
  \institution{University of Campinas}
  \country{Brazil}
}
\date{February 2023}

\copyrightyear{2023}
\acmYear{2023}

\setcopyright{acmcopyright}\acmConference[]{}
\acmBooktitle{}
\acmPrice{}
\acmDOI{}
\acmISBN{}

\begin{abstract}

Recent work has explored Large Language Models (LLMs) to overcome the lack of training data for Information Retrieval (IR) tasks. The generalization abilities of these models have enabled the creation of synthetic in-domain data by providing instructions and a few examples on a prompt.
InPars~\cite{10.1145/3477495.3531863} and Promptagator~\cite{https://doi.org/10.48550/arxiv.2209.11755} have pioneered this approach and both methods have demonstrated the potential of using LLMs as synthetic data generators for IR tasks.
This makes them an attractive solution for IR tasks that suffer from a lack of annotated data.
However, the reproducibility of these methods was limited, because InPars' training scripts are based on TPUs -- which are not widely accessible -- and because the code for Promptagator was not released and its proprietary LLM is not publicly accessible.
To fully realize the potential of these methods and make their impact more widespread in the research community, the resources need to be accessible and easy to reproduce by researchers and practitioners.
Our main contribution is a unified toolkit for end-to-end reproducible synthetic data generation research, which includes generation, filtering, training and evaluation. Additionally, we provide an interface to IR libraries widely used by the community and support for GPU.
Our toolkit not only reproduces the InPars method and partially reproduces Promptagator, but also provides a plug-and-play functionality allowing the use of different LLMs, exploring filtering methods and finetuning various reranker models on the generated data. We also made available all the synthetic data generated in this work for the 18 different datasets in the BEIR benchmark which took more than 2,000 GPU hours to be generated as well as the reranker models finetuned on the synthetic data. Code and data are available at \url{https://github.com/zetaalphavector/InPars}

\end{abstract}

\maketitle

\section{Introduction}
Effective neural Information Retrieval (IR) models often require a large amount of labeled training data. However, obtaining human labeled data is costly and many publicly available benchmarks contain few or no training examples~\cite{thakur2021beir}. In these cases, the common approach is to train a model on a large dataset, such as
MS MARCO~\cite{DBLP:journals/corr/NguyenRSGTMD16} and Natural Questions~\cite{kwiatkowski-etal-2019-natural}, and use it in a zero-shot transfer learning scenario~\cite{Rosa_2022}.
Nonetheless, models trained on these datasets face challenges to generalize to the variety of tasks and specific domains available in the real world.
Thus, the recently proposed InPars~\cite{10.1145/3477495.3531863} and Promptagator~\cite{https://doi.org/10.48550/arxiv.2209.11755} methods, along with their extensions InPars-v2~\cite{https://doi.org/10.48550/arxiv.2301.01820} and InPars-Light~\cite{https://doi.org/10.48550/arxiv.2301.02998}, have explored Large Language Models (LLMs) to generate synthetic data and have demonstrated their effectiveness. These methods not only outperform models that are finetuned on extensively labeled datasets but have also shown to be more adaptable to different tasks.

These methods propose the generation of synthetic in-domain training data by exploring the few-shot learning abilities of LLMs, prompting them with a brief description of the task and a small number of in-domain examples. InPars uses a static prompt that include examples collected from the MS MARCO dataset, whereas Promptagator uses dynamic prompts that include domain and task-specific examples sampled from the target dataset. Another key difference of these methods lies in the filtering of generated data. While InPars uses the sequence probability given by the LLM at generation time, Promptagator uses a consistency filtering with a model trained on the generated data. Similarly, InPars-v2 extends the pipeline by using a pre-trained reranker model to filter the examples. InPars-Light goes further in the efficiency direction by using lightweight models and showing that they are competitive with larger models.

These methods have proven to be effective, representing the state of the art in the BEIR benchmark~\cite{thakur2021beir}. However, reproducing such pipelines can still be a challenging task; researchers need to handle different codebases in addition to having access to a specific computational infrastructure. Most of the time, such components are not well integrated, making it difficult for researchers and practitioners to use them effectively. In this work, we bring all these components together, making it possible to experiment with InPars, Promptagator, and their variants, as well as to try new approaches using different LLMs, prompting approaches and datasets. We believe that making these resources available to allow reproducible work in the field of IR is crucial for several reasons. First, reproducibility is a key component of scientific research, as it allows other researchers to confirm and build upon the findings of a study. 
Second, the reproduction of LLM related studies is often costly, and making the models and generated data available provides a valuable resource for the community. 
We summarize our contributions as follows:

\begin{enumerate}[label=(\alph*)]
    \item We provide an extensive guideline for reproducing InPars and InPars-v2 for datasets on the BEIR benchmark on GPU. For Promptagator, we provide an implementation for reproducing the synthetic queries generation step with the dynamic prompt construction originally proposed by the authors.
    \item We also provide support for using different data sources: Pyserini's~\cite{Lin_etal_SIGIR2021_Pyserini} pre-built indexes for the BEIR datasets, and \texttt{ir\textunderscore datasets}~\cite{macavaney:sigir2021-irds} library, which contains multiple IR datasets.
    \item Lastly, we make available all the synthetic data generated in this reproduction study, along with the prompts and the finetuned reranker models.
\end{enumerate}

\section{Methods}

In this section, we describe the main methods reproduced in this paper and highlight the differences in their data generation pipelines.

\subsection{InPars}
The InPars method, currently available in two different versions, explores the few-shot learning abilities of LLMs to generate synthetic training data for IR tasks, by using a prompt template that instructs the LLM on how to generate the synthetic data. The prompt $t||d$ is the concatenation of a prefix $t$ and a document $d$, where the prefix $t$ consists of $N$ pairs of documents and their relevant queries, i.e., $t=\{(q_1^*,d_1^*), ..., (q_N^*,d_N^*)\}$. The prompt $t||d$ is fed to a language model $G$ that generates a question $q$ that is likely to be relevant to $d$. The resulting pair $(q, d)$ forms a positive training example that is later used to finetune a retrieval model. The original InPars work uses a GPT-3 LLM as the synthetic data generator, while InPars-v2 replaced the LLM with GPT-J~\cite{gpt-j}. These models, trained on massive amounts of text data, have shown impressive abilities in generating human-like text, answering questions, translating languages, and even creating original content. GPT-J is an open-source 6B parameters transformer model trained using 402 billion tokens from the Pile~\cite{DBLP:journals/corr/abs-2101-00027}, an 800 GB English dataset. When generating the synthetic queries, a greedy decoding strategy was used.

%

InPars proposes two different prompts. The first one, named ``Vanilla'' prompt, uses three fixed pairs of examples of document and relevant query, that were randomly collected from the MS MARCO training dataset. The second prompt template, referred to as ``Guided by Bad Questions'' (GBQ) uses the same examples from the first prompt, but it labels the original questions from the MS MARCO dataset as ``bad'' questions. The ``good'' questions were manually created and are more elaborated. The intention is to encourage the LLM to produce more informative questions, where the full context of the document contributes to the answers. 

InPars generates 100K pairs of positive training examples using documents randomly sampled from a collection $D$. The prefix $t$ is always the same regardless of the input document $d$. After generating the synthetic data, a filtering step is proposed, to select the top $K$ pairs with respect to the following (log) probability:
\begin{equation}
    p_{q} = \frac{1}{|q|} \sum_{i=1}^{|q|} \log p(q_i|t,d,q_{<i}),
\end{equation}
\noindent where $p(q_i|t,d,q_{<i})$ is the probability assigned by $G$ when autoregressively generating the $i$-th token of $q$, and $q_{<i}$ are the tokens generated in the previous decoding steps.
This score is used to filter the top $K=10,000$ pairs of document and synthetic queries to be used as finetuning data.
This filtering improves the quality of the training data. Without it, i.e., using the full set of 100K synthetic queries to finetune a reranker model resulted in a drop of 4 MMR@10 points on MS MARCO.

The filtering approach was improved on InPars-v2, where a pretrained reranker model is used to filter the synthetic queries for the training step. A monoT5-3B reranker model finetuned for one epoch on the MS MARCO dataset is used to estimate a relevance score for each synthetic query generated by the LLM and the document that was used to generate it. After computing the score for each one of the 100,000 pairs of synthetic queries and documents, only the $K=10,000$ highest scores are kept as finetuning data.

These filtered queries are used to train a monoT5~\cite{nogueira2020document} reranker, an adapted version of T5~\cite{DBLP:journals/corr/abs-1910-10683} model for text ranking tasks. The filtered queries are used as positive examples, while negative examples are mined from BM25 candidates. Two models with 220M and 3B parameters were trained for one epoch over the 20,000 query-document pairs. The trained model is subsequently used to rerank the initial BM25 retrievals. This approach employs a two-stage retrieval pipeline. Firstly, BM25 retrieves the top 1,000 documents per query. Secondly, the trained model reranks the list by assigning a relevance score for each pair of query and document.


\subsection{Promptagator}
The Promptagator method also creates synthetic training data for IR tasks by exploiting the few-shot abilities of a 137B-parameter LLM. Differently from InPars, a specific prompt is created for each dataset using in-domain examples.
%
By creating a specific prompt template for each dataset, the prefixes used were selected according to the dataset description. Using the ArguAna dataset prompt as an example, the model is prompted with a prefix ``\texttt{Argument:}'' which indicates the document from the dataset, followed by a prefix ``\texttt{Counter Argument:}'', marking the question related to the document. This way, the prefixes resemble a better description of the datasets while instructing the LLM  to generate a query for that specific task and document. Moreover, in the few-shot scenario, they use from 2 to 8 relevant query-document examples to create the prompt, sampled from the development set when it is available or, if not, from the test set. 

Promptagator generates synthetic queries using a sampling decoding algorithm with a temperature of 0.7. For each dataset, they generate 8 synthetic queries for each document from a randomly sampled set of 1 million documents. FLAN~\cite{wei2022finetuned} is used as the generator, which is a proprietary LLM that was pretrained on a multiple tasks using instructions.
To ensure that only high-quality synthetic questions are generated, the authors propose a filtering step based on consistency filtering. They train a retriever model using the same synthetic data that needs to be filtered to predict the most relevant passage for a given query. The retriever model keeps only queries that, when fed to the model, return the document that originated it among its top K results. The authors observed that setting K to 1 leads to better results when using the MS MARCO dataset as a validation set.
The authors suggest that this filtering strategy removes low-quality synthetic questions and improves performance on 8 out of the 11 datasets that were evaluated.

In the final step, the Promptagator method finetunes two different models using the synthetic data. The first is a bi-encoder based on the GTR~\cite{DBLP:journals/corr/abs-2112-07899} architecture with 110M parameters. The second is a cross-encoder with the same number of parameters, which reranks the top 200 candidates retrieved by the bi-encoder model. 

\section{Experimental Setup}

In this section, we describe the process of using the toolkit provided in this work. Firstly, we outline the steps for generating synthetic data. Next, we discuss the process of filtering the generated data to remove possibly irrelevant instances. Then, we describe how to build the training set using the filtered positive examples and mining the negatives. After that, we provide details on how to use the synthetic data to train a reranker. Finally, we describe the process of reranking and evaluating the trained model. By following the guidelines in this section, researchers and practitioners are able to leverage the provided resources effectively and reproduce the InPars method, as well as partially reproduce the Promptagator method and extend to new pipelines.

\subsection{Commands}
To begin, the synthetic data generation step is done using the \texttt{inpars.generate} command-line as follows: 
\begin{lstlisting}[language=bash]
  $ python -m inpars.generate \
        --prompt="inpars" \
        --dataset="trec-covid" \
        --dataset_source="ir_datasets" \
        --base_model="EleutherAI/gpt-j-6B" \
        --output="trec-covid-queries.jsonl" 
\end{lstlisting}
Diving into the required arguments, thwe first need to define the \texttt{prompt}, which supports four different options for the prompt template to be selected: \texttt{inpars}, \texttt{inpars-gbq}, \texttt{promptagator} and \texttt{custom}. This argument will define which prompt template to be used during the generation step. We provide both ``Vanilla'' and ``GBQ'' prompts templates used by InPars, with ``Vanilla'' as a default. 
The \texttt{promptagator} prompt template uses a specific template for each dataset and dynamically selects random pairs of query and relevant document to be used as prompt examples. The  \texttt{n\_fewshot\_examples} argument specifies the number of examples that will be used in the prompt in this case -- with a default of 3 examples.
We randomly select labeled examples from the training set of each dataset when training data is available. If there is no training set, we use the development set as our source and, as a last resort, when there is no training or development set, we use the test set for creating the prompt examples. This approach is slight different from the one proposed by Promptagator, that collects examples only from the development or test set. Once the examples are collected, for each document that requires a synthetic question, the prompt is built dynamically. This means that the prompt examples are randomly ordered for each document. 
To ensure a fair evaluation, the queries and documents used as few-shot examples that were extracted from the development or test sets are discarded from the evaluation metrics.

The next arguments are the \texttt{dataset} and \texttt{dataset\_source}, which specify dataset to generate synthetic queries for and the source from which to load it. In line with InPars and Promptagator, we support the datasets from the BEIR~\cite{thakur2021beir} benchmark. BEIR is a widely used evaluation framework in the IR domain. It aims to provide a comprehensive evaluation benchmark on a variety of IR tasks, with a particular emphasis on zero-shot evaluation. 

The \texttt{dataset\_source} argument is designed to integrate with two widely used dataset interfaces in the IR community: Pyserini~\cite{Lin_etal_SIGIR2021_Pyserini}, a toolkit for conducting reproducible IR research with sparse and dense representations, and \texttt{ir\_datasets}~\cite{macavaney:sigir2021-irds}, a commonplace for several IR \textit{ad-hoc} ranking benchmarks. Furthermore, it is also possible to indicate a local file as the document and query collection. By default, we use the \texttt{ir\_datasets} as the source, but both sources include all publicly available BEIR datasets.

The \texttt{base\_model} argument determines the LLM that will be used to generate the synthetic queries. By default, our toolkit uses the GPT-J~\cite{mesh-transformer-jax} model available in the Hugging Face Hub~\footnote{\url{https://huggingface.co/EleutherAI/gpt-j-6B}}, but it can support any generative model available from Hugging Face. Lastly, the \texttt{output} argument specifies the name of an output file to save the synthetic data. The output file will be a JSON format file, which will contain one query per line. This file will also include additional information related to the synthetic generation step, such as the log probabilities assigned to each token by the LLM during query generation, the prompt text fed to the LLM, and the document for which the query was generated.

Additional arguments related to the LLM, such as the maximum length of input and output or batch size, can also be set through the command-line arguments.

Once the synthetic data has been generated, we move on to the filtering stage. We provide two different filtering strategies, and the command to filter the synthetic queries is:

\begin{lstlisting}[language=bash]
  $ python -m inpars.filter \
        --input="trec-covid-queries.jsonl" \
        --dataset="trec-covid" \
        --filter_strategy="scores" \
        --keep_top_k="10_000" \
        --output="trec-covid-queries-filtered.jsonl"
\end{lstlisting}

Initially, before applying the filtering strategy, we keep only synthetic queries that meet some conditions. These conditions require the token count to fall within a specified range of minimum and maximum amount, defined by the arguments \texttt{min\_tokens} and \texttt{max\_tokens}. This is done to remove possible noisy synthetic queries. The \texttt{skip\_questions\_copied\_from\_context} optional argument removes synthetic queries in which a part of the document used for generation was copied to the query.

The first argument required by the filtering command-line is the \texttt{input}, which refers to the file containing the synthetic queries generated in the previous step to be filtered. The \texttt{dataset} argument indicates which dataset the queries belong to. The \texttt{filter\_strategy} specifies the filtering strategy to be used. The default filtering strategy, introduced by InPars-v1, is called \texttt{scores} and is based on a mean value computed from LLM's tokens probabilities. The synthetic queries list is then sorted in descending order, and only the top-K values are retained. The \texttt{keep\_top\_k} argument defines the value of $k$, with a default value of 10,000.

The second filtering strategy, which was introduced by InPars-v2, is called reranker. This strategy employs a pretrained reranker model to filter the synthetic queries by computing a relevancy score for each synthetic query-document pair. The scores list is sorted in descending order and only the top-k pairs with the highest scores are kept as positives query-document pairs to be used during training. Finally, the output file must be specified in \texttt{output} to indicate where the filtered synthetic queries will be saved.

The filtering strategy proposed by Promptagator is not currently supported because its more elaborated and seem to require more computational resources: A bi-encoder is initially finetuned on 1 million synthetic examples and then used in the filtering step by retaining only the examples that correctly retrieve the source document. This is a costly procedure that has been postponed for future work.

The third stage of the pipeline involves mining negative examples for model training. In this stage, negative examples are mined by using the filtered synthetic queries to search for candidate documents. We followed the approach outlined in InPars, using BM25 to retrieve 1,000 candidate documents from the target collection. From this set, a random document is selected as the negative example. If the candidate document is the same one used during the synthetic generation step, the example is discarded, and a new one is sampled. The following command-line is used to execute this step:
\begin{lstlisting}[language=bash]
  $ python -m inpars.generate_triples \
        --input="trec-covid-queries-filtered.jsonl" \
        --dataset="trec-covid" \
        --output="trec-covid-triples.tsv"
\end{lstlisting}
The \texttt{input} argument expects a file containing the previously filtered synthetic queries, as well as the \texttt{dataset} identification. The result is a tuple $(q, d^+, d^-)$, where $q$ and $d^+$ are fixed (the synthetic query and the source document) and $d^-$ represents the negative example sampled from BM25 candidates. The document collection is indexed using Pyserini, and all BEIR benchmark datasets are already available as pre-built indexes. 

Once the synthetic training data is obtained, we proceed to the training step. We support the finetuning of a monoT5 reranker model, which is the same model used in InPars, as the final stage of the multi-stage retrieval pipeline. 
To finetune the reranker using the synthetic data, the command-line is:
\begin{lstlisting}[language=bash]
  $ python -m inpars.train \
        --triples="trec-covid-triples.tsv" \
        --base_model="castorini/monot5-3b-msmarco-10k" \
        --output_dir="./reranker/" \
        --max_steps="156"
\end{lstlisting}
The \texttt{triples} argument specifies the file containing the training tuples obtained in the previous step, where every line consists of a triple comprising a query, a positive document, and a negative document. The \texttt{base\_model} argument indicates the model to be finetuned -- e.g., an original T5 model or a pre-trained monoT5. In all our experiments, we used the \texttt{castorini/monot5-3b-msmarco-10k}~\footnote{\url{https://huggingface.co/castorini/monot5-3b-msmarco-10k}} as our initial base model. The \texttt{output} argument specifies the path where the finetuned model should be saved. Our reranker models were trained for 156 steps, equivalent to one epoch over the query-relevant document pairs. In contrast to the InPars script that relies on TPUs, our training script supports GPUs. We conducted all experiments using a NVIDIA A100 80 GB GPU, and training the model for 156 steps took approximately 30 minutes.

\begin{table*}[h]
    \centering
    \begin{tabular}{llccccccccccc|c}
    \toprule
         & & covid & nfc & hotpot & fiqa & arg & touché & dbp & scidocs & fever & climate & scifact & AVG \\
         \midrule
         \midrule
         (1a) & BM25-flat       & 0.594 & 0.321 & 0.633 & 0.236 & 0.397 & 0.442 & 0.318 & 0.149 & 0.651 & 0.165 & 0.678 & 0.416 \\
         (1b) & BM25-multi    & 0.656 & 0.325 & 0.603 & 0.236 & 0.414 & 0.367 & 0.313 & 0.158 & 0.753 & 0.213 & 0.665 & 0.427 \\
         \midrule
         (2) & monoT5 (3B)~\cite{nogueira2020document}   & 0.795 & 0.384 & 0.759 & 0.514 & 0.288 & 0.200 & 0.478 & 0.197 & 0.850 & 0.280 & 0.777 & 0.511\\
         \midrule
         (3a) & InPars-v1  & 0.846 & 0.385 & 0.790 & 0.492 & 0.371 & 0.260 & 0.494 & 0.206 & 0.852 & 0.287 & 0.774 & 0.523\\
         (3b) & InPars-v2  & 0.846 & 0.385 & 0.791 & 0.509 & 0.369 & 0.291 & 0.498 & 0.208 & 0.872 & 0.323 & 0.774 & 0.533\\
         \midrule
         (4a) & Promptagator~\cite{https://doi.org/10.48550/arxiv.2209.11755} & 0.762 & 0.370 & 0.736 & 0.494 & 0.630 & 0.381 & 0.434 & 0.201 & 0.866 & 0.203 & 0.731 & 0.528 \\
         (4b) & InPars-v1$^\dagger$ & 0.762 & 0.364 & 0.606 & 0.487 & 0.506 & 0.262 & 0.490 & 0.175 & 0.858 & 0.307 & 0.790 & 0.519\\
         (4c) & InPars-v2$^\dagger$ & 0.832 & 0.357 & 0.790 & 0.487 & 0.529 & 0.327 & 0.394 & 0.192 & 0.857 & 0.139 & 0.790 & 0.517 \\
         \midrule
         (5) & RankT5 & 0.823 & 0.399 & 0.753 & 0.493 & 0.406 & 0.486 & 0.459 & 0.191 & 0.848 & 0.275 & 0.760 & 0.535\\
        \bottomrule
    \end{tabular}
    \caption{Results from monoT5 reranker finetuned using synthetic data generated from the InPars and the Promptagator prompts. $^\dagger$  indicates that Promptagator prompt templates were used to generate the synthetic data.}
    \label{tab:tab_overview}
\end{table*}

Once the model has been trained, the next stage uses it to rerank a dataset. In this stage we support all BEIR datasets, as well as any custom local datasets. The command-line to rerank is:
\begin{lstlisting}[language=bash]
  $ python -m inpars.rerank \
        --model="./reranker/" \
        --dataset="trec-covid" \
        --output_run="trec-covid-run.txt"
\end{lstlisting}
The first argument is the \texttt{model}, which specifies the trained model to be used for reranking. The \texttt{dataset} argument indicates one of the BEIR datasets to load the documents and queries, as well as the initial run to be reranked. We are using the BEIR runs, created using BM25, as initial run for all the datasets. However, it is possible to provide an initial run from a local file in the TREC format using the \texttt{initial\_run} argument. The reranker model will compute a relevancy score for each query and the candidates documents from the initial run. The output will consist of a reranked run, which will be saved in the location indicated by the \texttt{output\_run} path.

Finally, to evaluate the reranked run, the following command-line is used:
\begin{lstlisting}[language=bash]
  $ python -m inpars.evaluate \
        --dataset="trec-covid" \
        --run="trec-covid-run.txt"
\end{lstlisting}
By providing the \texttt{dataset} and \texttt{run} to be evaluated, our script computes the metrics like recall and nDCG, in addition to other metrics computed by the TREC evaluation script. 

\section{Results}

This section presents and discusses the results obtained by reproducing the methods using our toolkit. Table~\ref{tab:tab_overview} presents a comparison between the baselines, the results reported by original methods and our reproductions. 
The first two rows (1a and 1b) represent the BM25 baselines. In BM25-flat, document titles and contents are concatenated and stored as a single field while BM25-multi stores them as separate fields. The top 1000 documents retrieved by BM25-flat are reranked by the models in rows (2), (3a), (3b), (4b), and (4c). Row (2) presents the result of monoT5-3B, which was finetuned on MS MARCO for one epoch, used in a zero-shot setup.
Rows (3a) and (3b) presents the reproductions of InPars-v1 and InPars-v2 pipelines, respectively. Row (4a) presents the reported result for Promptagator.
The results in rows (4b) and (4c) illustrate the impact of using the Promptagator prompt with InPars pipelines. Comparing these results to those of InPars v1 and v2 (rows (3a) and (3b)), the results produced by the Promptagator prompt are either equal or slightly lower than those obtained through the InPars prompt, except for the ArguAna, Touché-2020 and SciFact datasets. Notably, for the ArguAna dataset, finetuning the reranker on the synthetic data generated by the Promptagator prompt resulted in an almost 14 nDCG@10 improvement compared to InPars' best result. These findings suggest that the Promptagator prompts are particularly effective in generating synthetic queries for the ArguAna and Touché-2020 datasets. Such datasets concentrate on argument retrieval, which is slightly different from other datasets in the BEIR benchmark. As a result, they gain advantage from using dataset-specific prompts. 
Also, a factor that probably limited InPars prompt performance on the ArguAna dataset reported in rows (3a) and (3b) is related to the query length. When generating the synthetic queries, InPars sets a maximum number of 64 tokens. As shown in Table~\ref{tab:tab_tokens}, the average number of words and tokens for queries across all datasets in the BEIR benchmark is below this value with the exception of the ArguAna dataset.

When examining the filtering strategy, the results from the InPars prompt indicate an average difference of 1 nDCG@10 point between the results displayed in rows (3a) and (3b). The improvements observed in the reranker strategy results are primarily driven by Touché-2020, FEVER and Climate-FEVER datasets. When using the Promptagator prompt, the filtering strategy appears to make a difference for certain datasets, as shown in rows (4b) and (4c). The reranking strategy appears to perform better for TREC-COVID, Touché-2020, and HotpotQA datasets. In particular, the HotpotQA dataset showed an improvement of more than 18 nDCG@10 points when compared to the scores strategy. On the other hand, the scores filtering strategy resulted in an improvement for the DBPedia and Climate-FEVER datasets, with gains of 9.6 and 16.8 nDCG@10 points, respectively. Despite the individual differences, the average results are very similar.

\begin{table}[h]
    \centering
    \begin{tabular}{l|c|c|c}
        \toprule
         & TPU & GPU & Diff \\
         \midrule
         TREC-COVID       & 0.846 & 0.851 & 0.005 \\
         BioASQ           & 0.595 & 0.629 & 0.034 \\
         NFCorpus         & 0.385 & 0.385 & 0.000 \\
         NQ               & 0.638 & 0.639 & 0.001 \\
         HotpotQA         & 0.791 & 0.711 & 0.080 \\
         FiQA-2018        & 0.509 & 0.506 & 0.003 \\
         Signal-1M (RT)   & 0.308 & 0.308 & 0.000 \\
         TREC-NEWS        & 0.490 & 0.490 & 0.000 \\
         Robust04         & 0.632 & 0.648 & 0.016 \\
         ArguAna          & 0.369 & 0.407 & 0.038 \\
         Touché-2020      & 0.291 & 0.287 & 0.004 \\
         CQADupstack      & 0.448 & 0.441 & 0.007 \\
         Quora            & 0.845 & 0.844 & 0.001 \\
         DBPedia          & 0.498 & 0.498 & 0.000 \\
         SCIDOCS          & 0.208 & 0.212 & 0.004 \\
         FEVER            & 0.872 & 0.869 & 0.003 \\
         Climate-FEVER    & 0.323 & 0.323 & 0.000 \\
         SciFact          & 0.774 & 0.775 & 0.001 \\
         \midrule
         AVG              & 0.545 & 0.545 & 0.010 \\
         \bottomrule
    \end{tabular}
    \caption{Reproduction of InPars-v2 on TPU and GPU. Numbers are nDCG@10. The same synthetic data was used to train each model on different devices.}
    \label{tab:tab_tpu_gpu}
\end{table}

Table~\ref{tab:tab_tpu_gpu} presents results comparing the performance on GPU (PyTorch~\cite{Paszke_PyTorch_An_Imperative_2019} and Transformers~\cite{Wolf_Transformers_State-of-the-Art_Natural_2020}) versus TPU (Mesh-TensorFlow~\cite{shazeer2018mesh}). As part of our work, we added GPU support to reproduce InPars results. This support covers the synthetic data generation, filtering, training, reranking and evaluating. We conducted an experiment to verify that running it on a GPU setup would produce the same results as running it on the TPU setup. For this, we trained monoT5-3B models following the InPars-v2 approach. Our analysis revealed that while there were minor variations in datasets such as TREC-COVID, BioASQ, Robust04 and ArguAna, the results remained exactly the same for NFCorpus, NQ, and FiQA-2018, regardless of the device used. For the majority of datasets, the variance in results between running on TPU and GPU is minimal when considering individual performance, as demonstrated in the "Diff" column on Table~\ref{tab:tab_tpu_gpu}. Furthermore, the average nDCG@10 remains consistent in both evaluation scenarios.

All experiments were conducted using an NVIDIA A100 80 GB GPU. Training monoT5-3B for 156 steps took about 30 minutes. Filtering 100K queries using a monoT5-3B model takes approximately 45 minutes. The duration of the evaluation step is determined by the number of queries that need to be reranked for each dataset, which can range from 50 queries for TREC-COVID to 13,145 queries for CQADupstack. The reranking of 1,000 candidate documents for a given query took a maximum of 30 seconds using the monoT5-3B reranker model.

Additionally, Table~\ref{tab:tab_tokens} shows statistics regarding the token count for each set of documents and queries in all datasets on the BEIR benchmark. 
\begin{table}[h]
    \centering
    \begin{tabular}{l|c|c|c|c}
        \toprule
         & \multicolumn{2}{c}{Documents} & \multicolumn{2}{c}{Queries} \\
         & Words & Tokens & Words & Tokens \\
         \midrule
         ArguAna          &  170 & 210 & 197 & 246 \\
         TREC-NEWS        &  655 & 918 &  11 &  14 \\
         Robust04         &  469 & 617 &  15 &  18 \\
         Touché-2020      &  293 & 370 &   6 &   7 \\
         NFCorpus         &  220 & 322 &   3 &   5 \\
         BioASQ           &  203 & 310 &   8 &  12 \\
         SciFact          &  201 & 303 &  12 &  19 \\
         TREC-COVID       &  197 & 289 &  10 &  15 \\
         SCIDOCS          &  169 & 220 &   9 &  14 \\
         CQADupStack      &  146 & 274 &   8 &  11 \\
         FiQA-2018        &  136 & 172 &  10 &  13 \\
         Climate-FEVER    &   99 & 123 &  20 &  25 \\
         FEVER            &   97 & 117 &   8 &  11 \\
         NQ               &   79 & 106 &   9 &  10 \\
         DBPedia          &   50 &  75 &   5 &   7 \\
         HotpotQA         &   46 &  69 &  15 &  20 \\
         Signal-1M (RT)   &   15 &  23 &   9 &  10 \\
         Quora            &   12 &  15 &   9 &  11 \\
         \bottomrule
    \end{tabular}
    \caption{Average number of words and tokens for each dataset in the BEIR benchmark. We use Python's \texttt{str.split()} for counting the number of words. For tokenizing the text, we use GPT-J tokenizer. }
    \label{tab:tab_tokens}
\end{table}
The ArguAna dataset is noteworthy for having significantly different query length compared to the other datasets. TREC-NEWS and Robust04 have the largest document lengths. This information is crucial to keep in mind when choosing documents to use as prompt examples. For instance, if we consider the GPT-J model, with a maximum sequence length of 2048 tokens, at most two average TREC-NEWS documents can fit into a prompt, without even accounting for the length of the queries.

\section{Conclusions}

We have introduced the InPars Toolkit, a codebase designed to generate synthetic data using LLMs in a reproducible manner for neural IR tasks. The toolkit comprises an end-to-end pipeline that encompasses data generation, training, reranking, and evaluating the trained models. Additionally, the codebase is integrated with two major libraries for commonly used datasets from the BEIR benchmark, and it supports both GPU and TPU training and inference. Our goal is to make research on these methods more accessible and to pave the way for this emerging research trend in the IR community.
Our experiments have demonstrated that training reranker models using synthetic data and evaluating them on GPU infrastructure yielded results comparable to those obtained when training on the TPU setup. Additionally, we have also made available all synthetic data generated for all BEIR datasets and the models finetuned on this data.

\section{Future Work}

Future work will focus on integrating a wider range of open-source LLMs, including instruction finetuned LLMs, with the aim of enhancing the generation process. Another area of further exploration is to experiment with different prompting techniques, such as chain-of-thought prompting, and prompting for retrieval explanations. Moreover, there are plans to incorporate consistency filtering and expand the filtering methods to completely reproduce Promptagator and lay the foundations for new research approaches in the field of synthetic data generation for IR.

\bibliographystyle{abbrv}
\bibliography{ref}
\end{document}